\documentclass[preprint,preprintnumbers,amsmath,amssymb]{revtex4}

\usepackage{graphicx,,bm}

\begin{document}
\title{Polysiloxane surfactants for the dispersion of carbon nanotubes in non-polar organic solvents}

\author{Y. Ji}
\author{Y. Y. Huang}
\author{A. R. Tajbakhsh}
\author{E. M. Terentjev}   \email{emt1000@cam.ac.uk}
\affiliation{Cavendish Laboratory, University of Cambridge, J.J. Thomson Avenue, Cambridge CB3 0HE, U.K.}

\begin{abstract}
\noindent We develop two new amphiphilic molecules that are shown
to act as efficient surfactants for carbon nanotubes in non-polar
organic
solvents. The active conjugated groups, which are highly attracted to
graphene nanotube surface, are based on pyrene and porphyrin. We
show that relatively short (C18) carbon tails are insufficient to
provide stabilization. As our ultimate aim is to disperse and
stabilize nanotubes in siloxane matrix (polymer and crosslinked
elastomer), both surfactant molecules were made with long siloxane
tails to facilitate solubility and steric stabilization. We show
that pyrene-siloxane surfactant is very effective in dispersing
multi-wall nanotubes, while the porphyrin-siloxane is making
single-wall nanotubes soluble, both in petroleum ether and in
siloxane matrix.
\end{abstract}

\maketitle

\section{Introduction}
With their unique mechanical and electrical properties, and surface
chemistry~\cite{Dresselhaus01_NTproperty,Banerjee05_CNTcovalentChemistry},
carbon nanotubes (CNTs) have attracted considerable interests in
the studies of their fundamental properties as well as the
adaption for a wide range of industrial applications. In many
situations a homogeneous dispersion of isolated CNTs in solution
is required. A bundled CNT system can lead to an uncontrolled
alteration in
fundamental attributes (e.g. a (n, n) SWNT bundle will display
pseudo-semiconducting rather than metallic
character~\cite{Delaney98_CNTRopeBrokenSym}), or result in poorer
performances (e.g. decrease in the effective stress transfer in
composites~\cite{Bokobza07_ElastomerReview,Shi04_dispersionEffectStrength}).
Generally, CNTs in big bundles or dense aggregates of other
morphology are not very different from ordinary carbon black; for
any advanced application one needs them well-separated in a
matrix. Unbundling and dispersion of pristine CNTs can be assisted
by the use of dispersants or surfactants through non-covalent
functionalization. Compared to covalent surface modification of CNTs,
non-covalent functionalization has the benefit of preserving the $\pi-\pi$
electronic structures of the outer CNT surfaces.
Although a fair amount of research has been
conducted on this topic, much of today's surfactants are based on
ionic~\cite{Chen05_PhrAquous,Islam03_SDS,Zheng03_DNAseparateSWNT}
or highly conjugated
structures~\cite{Zou08_BlcokCopolymer,Dalton00_ConjugateDisper,Cheng08_FluoreneConjugated},
which are mostly suitable for dispersion in aqueous solutions or
selected polar solvents (e.g. dimethylformamide,
N-methylpyrrolidone, terahydrofluran). The development of a
surfactant which can enhance the CNT solubility in generic
non-polar organic solvents is of great technical importance
especially in the composite fabrication.

An ideal dispersant molecule for CNTs should have an
``amphiphilic''  structure with an active group attracted towards
the graphene wall of a nanotube, and a flexible moiety matching the chemistry of
the solubilizing medium and having a carefully chosen size to
ensure proportional coverage of the CNT surface. Pyrene and
porphyrin are two of the most
studied~\cite{Ehli06_PyrenePorphyrin,Satake05_PhrWrap,Chen01_ProteinImmobil,Li03_PhrSemiCondSWNT}
functional groups highly interacting with CNTs. Therefore, the
simplest concept to design our ideal dispersant would be to link
one of these functional groups to a simple saturated alkyl chain.
Such systems based on pyrene/porphyrin derivatives grafted to
SWNTs have been studied
before~\cite{Satake05_PhrWrap,Tanigaki07_PyreneCNT}, however, they
did not show remarkable enhancement in CNT solubility in organic solvents. Based on
this past experience, we suggest that for stabilizing CNTs in generic
solvents, some strict criteria exist for both the CNT-philic and
solvent-philic parts of the surfactant.

Although this paper focuses on the CNT dispersion in a generic
non-polar solvent, our ultimate aim and interest lie in the
nanotube dispersion in silicone matrix. There are many important
reasons to produce polymer nanocomposites based on siloxane
elastomers, which are highly elastic, chemically stable,
insulating materials with a very low glass transition temperature.
This explains our use of siloxane moieties in some of our
dispersants. However, in order to mix CNTs into siloxane polymers
one first needs to assure their miscibility with low-molecular
weight solvents, which is what we now proceed to discuss.

We systematically investigated criteria for an effective
surfactant to disperse CNTs in non-polar solvents, choosing
petroleum ether (PET) as a model. A combination of different
active centers (pyrene or porphyrin) and chain attachments (alkyl
or siloxane chains) of varying lengths were tested, leading to the
successful development of dispersants for SWNTs and MWNTs. Since
the nanohybrids formed by integration of pyrene/porhyrin and CNTs
have seen promising potential in photovoltaic
applications~\cite{Hasobe06_PhrSWNTSolarCell}, there is an
current excitement in these systems; our findings are
expected to greatly enhance the processability of these hybrids
into useful matrices such as generic silicone elastomers.

\section{Experimental}
In all cases, the organic synthesis has been an extremely simple
one-step process which makes these new materials attractive for
practical applications.

\subsection{Pyrene surfactant: mPSi$_{70}$}
The core reactants for the synthesis of mono-pyrene siloxane (mPSi$_{70}$) are:
polydimethylsiloxane (PDMS) (bis(hydroxyalkyl) terminated,
Mn=5600~{g/mol}, from Aldrich, with an estimated number of
siloxane units $n= 70$) and 1-pyrenebutyric acid ($M_{w}\sim
288~\hbox{g/mol}$, from Aldrich), added in a $1:2$ ratio. The
pyrenebutyic acid powder was first dissolved in dichloromethane
(DCM), then mixed with PDMS in the presence of excess
N,N-diisopropylcarbondiimide. The reaction mixture was stirred
overnight and the solid by-product filtered; the filtrate was then
evaporated under reduced pressure to give an oil, which was
precipitated from ethanol. The IR spectrum showed a new absorption peak
 at $1740~\hbox{cm}^{-1}$ corresponding to the ester group of
the butyric acid, thus implying the successful formation of
mPSi$_{70}$. The ideal structure of our new surfactant, which we
designate as mono-pyrene-siloxane (mPSi$_{70}$), is shown
schematically in Table~\ref{Table_sum}.
The NMR spectrum of the
purified product indicated a 9:380 ratio between the protons on
the aromatic ring of pyrene ($\sim 7.8-8.3$~ppm) and those on the
methyl groups of PDMS ($\sim 0.1$~ppm). Each siloxane chain of $70$ monomers
 would have 70*6=420 protons, and there are 9 protons on each pyrene moiety.
 If we assume the fraction of mono-substituted pyrene-siloxane is $x$, and
 thus $(1-x)$ is the fraction of di-pyrene-siloxane, then the following relation holds:
$[9x + 18(1-x)] : 420 = 9 : 380$. This gives $x = 0.89$, that is, around
10\% of the substitution is to form doubly substituted di-pyrene siloxane (dPSi$_{70}$)
since we have essentially no un-reacted siloxane. The rest (the majority $90\%$) is the
mono-pyrene siloxane mPSi$_{70}$. The IR and NMR spectra of the mPSi$_70$
product are available in Supporting information.

Allowing the small proportion of di-pyrene siloxane was intentional (and is
the reason for the initial 1:2 ratio of the reactants). First of all, aiming
for the strict mono-substitution and have the siloxane/pyrene reacting groups
in the 1:1 molar ratio would significantly reduce the reaction yield and make
purification more difficult. The excess of un-reacted pyrene is easily washed
off by hot ethanol, while the un-reacted PDMS (which would be left if a 1:1 ratio
was used) is really hard to separate from pyrene-PDMS. As it was, the reaction
yield was above 90\% after one day.  Secondly,
and perhaps more importantly because one might find other ways to improve
the yield and purification, we believe
that a fraction of di-functional molecules enhance the stability
of CNTs at high concentrations by bridging between neighboring
tubes. This is no more than an opinion and the reason it is rather subtle:
it is based on our earlier experience in rheological studies of dispersed CNTs~\cite{Huang06_PRB,Ahir07_Networks} which indicated that a swollen
gel network of effectively crosslinked CNTs has the effect of preventing
re-aggregation into compact bundles.

\subsection{Porphyrin surfactant: PhrSi$_{60}$}
PDMS (mono(hydroxyalkyl) terminated, Mn=4670~g/mol, from Aldrich, with an
estimated number of siloxane units $n=$60) and 5,10,15,20-Tetrakis
(4-carboxyphenyl) porphyrin (TPCC, from Aldrich), were added in a 1:16 ratio.
The TPCC powder was first dissolved in DCM, then mixed with PDMS in the
presence of excess N,N-diisopropylcarbondiimide. The reaction mixture was
stirred for two days and the solid by-product filtered; the filtrate was
then evaporated under reduced pressure to give an oil, which was
precipitated in ethanol. The oil was re-dissolved in isopropanol, followed by the
addition of appropriate amount of acetone. Subsequently, the mixture was
cooled in a freezer, and the
dark red layer of the separated product collected. The precipitation procedure
described above was repeated several times until there was only one peak remained
in gel-permeation chromatography (GPC) analysis. Similar to the mPSi,
the IR spectrum (see Supporting Information) showed an
absorption peak at 1722~cm$^{-1}$ corresponding to the attachment of ester groups to
the porphyrin, implying the successful attachment of siloxane to porphirin. NMR
spectrum of the purified product indicated that all of the four carboxyl
acid groups of porphyrin have been reacted with PDMS (details in Supporting Information).

\begin{table}
\centering
\resizebox{0.8\textwidth}{!}{\includegraphics{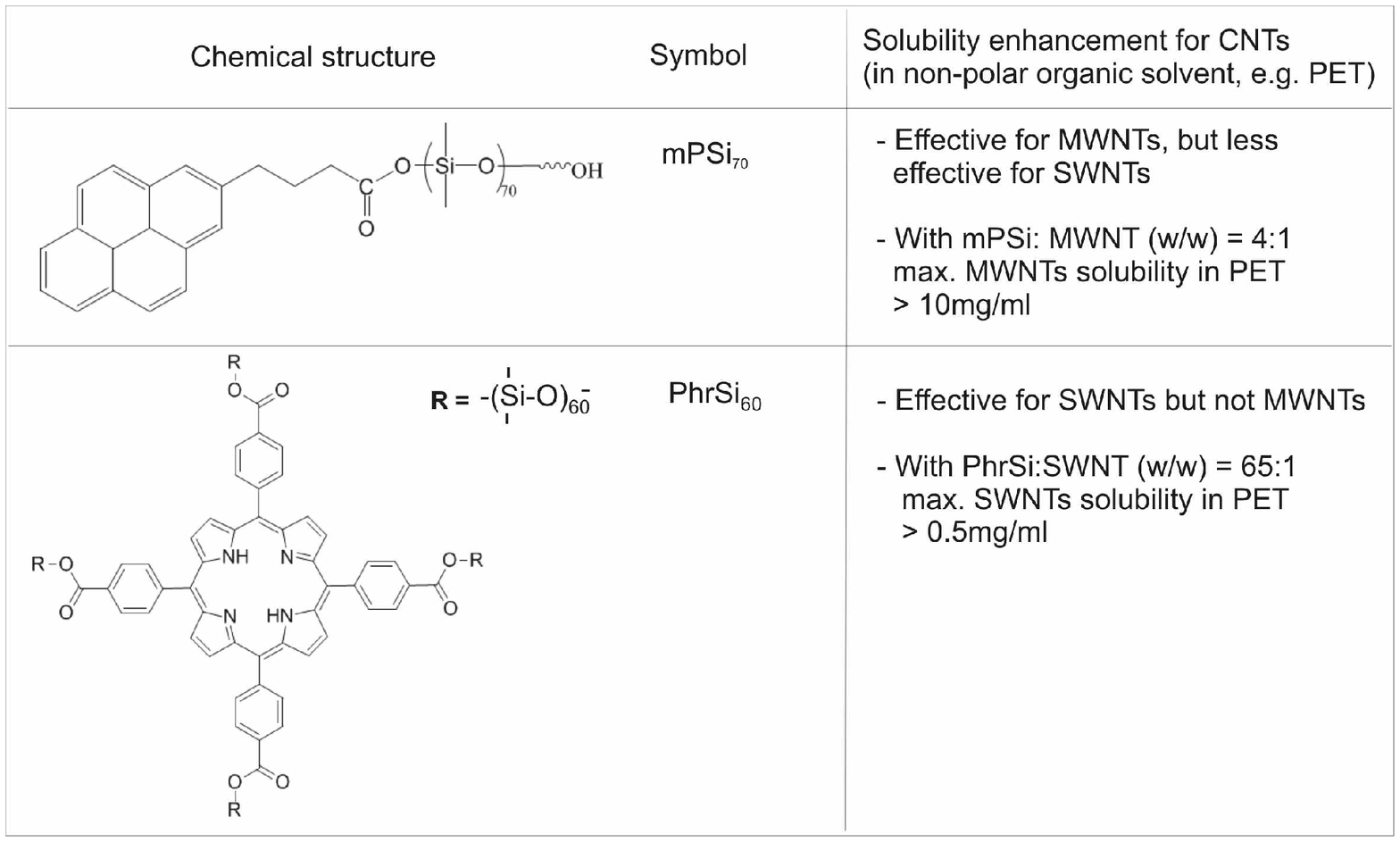}}
\caption{Summary of the mPSi$_{70}$ and PhrSi$_{60}$ surfactant structures and their effectiveness} \label{Table_sum}
\end{table}

\subsubsection*{$\bullet$ \ Short chain porphyrin surfactants: PhrC$_{10}$ and PhrC$_{18}$}
The synthesis process for 5,10,15,20-Tetrakis (n-decoxycarbonyl) porphyrin (PhrC$_{10}$), and 5,10,15,20-Tetrakis
(n-octadecoxycarbonyl) porphyrin (PhrC$_{18}$) is similar to
that of PhrSi$_{60}$ except that the reactants were different. For PhrC$_{10}$,
the reactants were Octadecan-1-ol (1.62~g, 5.99~mmol) and TPCC (0.10~g, 0.12~mmol).
For PhrC$_{18}$, the reactants are Decyl alcohol-1-ol (0.95~g, 6~mmol) and TPCC (0.38~g, 0.48~mmol).
However, these two surfactants did not give rise to solubility enhancement
for either MWNTs and SWNTs in PET. Therefore, they were not investigated in
detail in the present manuscript.

\subsection{Dispersion of carbon nanotubes}~\label{CNTDispersion}
Our MWNTs were obtained from Nanostructured $\&$ Amorphous
Materials, Inc (CVD, purity 95+\%, diameter 60-100 nm, as-produced
length 5-15~$\mu$m), with no functionalization. SWNTs were
purchased from Carbon Solutions, Inc., under the  P2-SWNT grade
(carbonaceous purity of 90\%, low functionality and low
chemical doping). These SWNTs were grown by electric arc
discharge, and consisted of, theoretically, metallic to
semiconducting nanotubes in the standard ratio of $1:2$. The SWNTs
were quoted to have bundle length between 500~nm to 1.5~$\mu$m,
and bundle diameter in the region of 4-5~nm. The Scanning electron micrographs
(SEM) of these two two commercial nanotubes sources are shown below
in Figs.~\ref{MWNT_son}(A) and~\ref{SW_soln}(A).
Their purity were further confirmed by thermogravimetric analysis
(TGA, results in Supporting Information).

Sonication was employed to assist the dispersion of nanotubes in
solution.  The effect of sonication on the structural changes of
nanomaterials has been recently reviewed in
\cite{Ahir08_CompositeReview}, which also includes the analysis of
CNT scission induced by ultrasonic energy. Even though in the
present study of miscibility we were not particularly concerned
about the breakage of nanotubes, due to the excessive power
density transmitted into the solution, the sonication conditions
were carefully controlled to ensure reproducibility. Ultrasonic
Processor (Cole Parmer, $750$~W model) with titanium micro-tip was used.
The sonication conditions were kept consistent by using standard
glass bottles and a fixed cell geometry. The sonication cell has a
cooled water bath which kept the ambient temperature of the bottle
at 12-16~$^{\circ}$C, and also prevented overheating of the
solution. The bath water was filled to a constant level throughout
the experiments,  ensuring that the location of micro-tip emitter
and the bottle were both centered within the cylindrically shaped
bath. The micro-tip was located at $0.5$~cm above the bottom of
the bottle and the glass bottle was suspended $\sim4.5$~cm above
the bath bottom. The following sonication parameter settings were
kept the same throughout all experiments: pulsar with 5\,s
on and 3\,s off, probe temperature at 15~$^{\circ}$C and
vibrational amplitude at $25$~\%. All sonications were performed
with a 1\,hr of actual pulsing time. No change in the chemistry of
surfactants were found after sonication as determined by
absorption spectroscopy (sonochemistry is a known issue in this
context
\cite{Soni_invasive_Leighton05,Review_Soni_PolymerChemistry06}). A
standard procedure of dispersion that we followed involved first
dissolving the $m$ mass of surfactant into $\sim$6~ml of PET
(petroleum ether, 40-60$^{\circ}$C). The solution is then
transferred to the holder which contains the $n$ mass of CNTs. The
$m:n$ ratios required for different surfactants are shown
in Table~\ref{Table_sum}. After soaking the CNTs in the surfactant
solution overnight, sonication was applied to debundle the nanotubes.

\subsection{Characterization and data analysis}
Thermogravimetric analysis (TGA) of the raw (as-received) CNTs was
performed on the TA Instruments Q500 device.
Infrared (IR) and Nuclear Magnetic Resonance (NMR) spectroscopy
were  employed to determine the nature of the reaction products.
IR spectra were obtained with Thermo Scientific Nicolet iS10
Spectrometer, and NMR were performed with Bruker Avance 500MHz NMR
Spectrometer.

Scanning electron micrographs (SEM) of all CNT samples were obtained
on FEI XL30-SFEG high-resolution scanning electron microscope. For the
non-surfactant-added reference (see Fig.~\ref{MWNT_son}(B)), sonicated MWNTs-PET solution
was directly dropped onto mica sheet and the PET was allowed to
evaporate in open air. On the other hand, SEM of the sonicated mPSi$_{70}$/MWNT (Fig.~\ref{MWNT_son}(C)) required more careful preparation.
The sample was prepared by first drop-casting the as-sonicated
mPSi$_{70}$/MWNT PET solution onto a mica sheet,
after which a few drops of toluene were used to dissolve and wash
off the excess mPSi$_{70}$, leaving the residue toluene evaporating in open air.

In order to examine the interaction between the active centers of
the surfactants (pyrene or porphyrin) and the nanotube surfaces,
absorption spectra of different CNT and surfactant solutions were
obtained by CARY 300Bio UV-Vis spectrometer.
Absorption peak analysis was performed by OriginPro 8 (V8 SR2,
OriginLab) through the ``Peak Analyzer'' function. Steady-state
fluorescence emission measurements were carried out at room
temperature using a CARY Eclipse fluorescence spectrophotometer
from Varian. For pyrene, the excitation wavelength was 300nm;
for porphyrin it was 420nm.

\section{Results}~\label{ResultsSec}
Unmodified pyrene, neutral porphyrin, and silicone are all soluble
in PET at room temperatures, thus one finds that the synthesized
surfactants could all be easily dissolved in PET. However, neither
pristine SWNTs or MWNTs demonstrated any solubility in PET. Fast
re-aggregation and precipitation of nanotubes was observed
immediately after sonication, leaving clear supernatant, e.g. Fig.~\ref{mP70Si_soln}(C).

What happens after adding the surfactants? Following the
dispersion  procedure described above, we
determined the effectiveness of each surfactant by observing the
stability of sonicated solutions with different CNT contents and
surfactant levels. If no apparent sediment appeared in solution
after three days' waiting, the surfactant/CNT combination was
considered to be adequate. Table~\ref{Table_sum} summarizes the
effectiveness of the different surfactants we synthesized.

First of all, no solubility enhancement was found for both SWNTs
and MWNTs when PhrC$_{10}$ and PhrC$_{18}$ were used. These two
surfactants both result in a pink color when dissolved in PET. Significant decoloration
was observed after the surfactant solution was sonicated with
SWNTs, but not MWNTs. Since absorption spectroscopy showed no
surfactant degradation under our sonication condition, the
decoloration phenomenon strongly suggests the adsorption of
porphyrin-derived surfactants by SWNTs. By comparing the UV/vis
absorption of the sonicated PhrC-PET solution and
the sonicated PhrC/SWNTs-PET solution with the same starting PhrC concentration,
one finds nearly complete disappearance of the absorption peaks due to PhrC.
This further implies the interaction between the porphyrin surfactants and SWNTs,
though no solubility enhancement has been observed.

In contrast to the porphyrin with short carbon chain attachments,
mPSi$_{70}$ is shown to be a highly effective surfactant for
MWNTs, whereas PhrSi$_{60}$ worked well with SWNTs. In the
following sections we will bring our focus to these two
surfactants, and then discuss the possible reasons for the
observed ``selective solubilization'' effect.

\subsection{mPSi$_{70}$}
A remarkable improvement in the solubility of MWNTs in PET was
obtained  with the addition of mPSi$_{70}$. For SWNTs, however,
only a stable suspension of small clusters could be achieved even
when high concentrations of mPSi$_{70}$ were tested. Although
these clusters appeared to be stable over a few weeks, mPSi$_{70}$
was not considered to be a good surfactant for SWNTs: our criteria
demanded full dispersion. Therefore, we would like to focus our
discussion on MWNTs in this section.

\begin{figure}
\centering
\resizebox{0.3\textwidth}{!}{\includegraphics{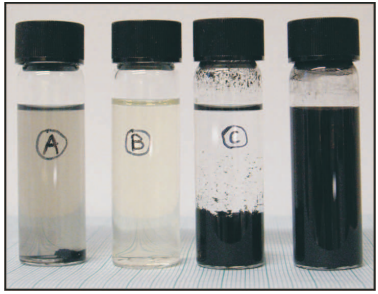}}
\caption{Bottled samples showing the effect of mPSi$_{70}$:
(A) partial solubilization of mPSi$_{70}$ soaked MWNTs in PET
without sonication; (B) light-yellow colored mPSi$_{70}$ in PET
reference solution; (C) sedimentation of 1~mg/mL MWNTs in PET seen
$10~$min after sonication; (D) stable 1~mg/mL mPSi$_{70}$/MWNTs in
PET solution seen 2 weeks after sonication. }
\label{mP70Si_soln}
\end{figure}

Figure~\ref{mP70Si_soln} compares the results of four samples that
underwent different treatments. The MWNTs in sample (A) had been
soaked in a few drops of pure mPSi$_{70}$ (which is an oily liquid
at room temperature) overnight, after which PET was added to the
cluster. By just gentle stirring, some tubes started to diffuse
and disperse into PET, resulting in a semi-transparent dark
supernatant which remained stable for more than 2 weeks,
contrasting in color with the pure surfactant solution shown in
sample (B). On the other hand, complete sedimentation was found
for MWNTs sonicated in pure PET, illustrated in sample (C). Sample
(D) presents a homogeneous mPSi$_{70}$/MWNT solution at a MWNTs
concentration of 1~mg/ml, with no visible sedimentation after 2
weeks' standing.

The attachment of mPSi$_{70}$ to MWNTs was strong after sonication
treatment, such that the dense sediment obtained after repeated
centrifugation (for 40~min at $8000$~rpm, which corresponds to
acceleration of $ \approx 5700\,\hbox{g}$ in our reactor geometry)
and dilution
can still
be easily re-dispersed. A convenient mPSi$_{70}$:MWNT ratio for
the formation of stable solutions was found to be $\sim 4:1$\,w/w
in PET, although a lower ratio of mPSi$_{70}$ to MWNT was also
possible. It was found that with the $4:1$ mPSi$_{70}$:MWNT
standard, over $10$~mg/ml ($\sim1$~wt\%) MWNTs in PET could be
stabilized, though sediment may appear after three days at very
high concentrations.  It is expected that this ratio would change
depending on the diameter of pristine MWNTs.

\begin{figure}[t]
\centering
\resizebox{0.9\textwidth}{!}{\includegraphics{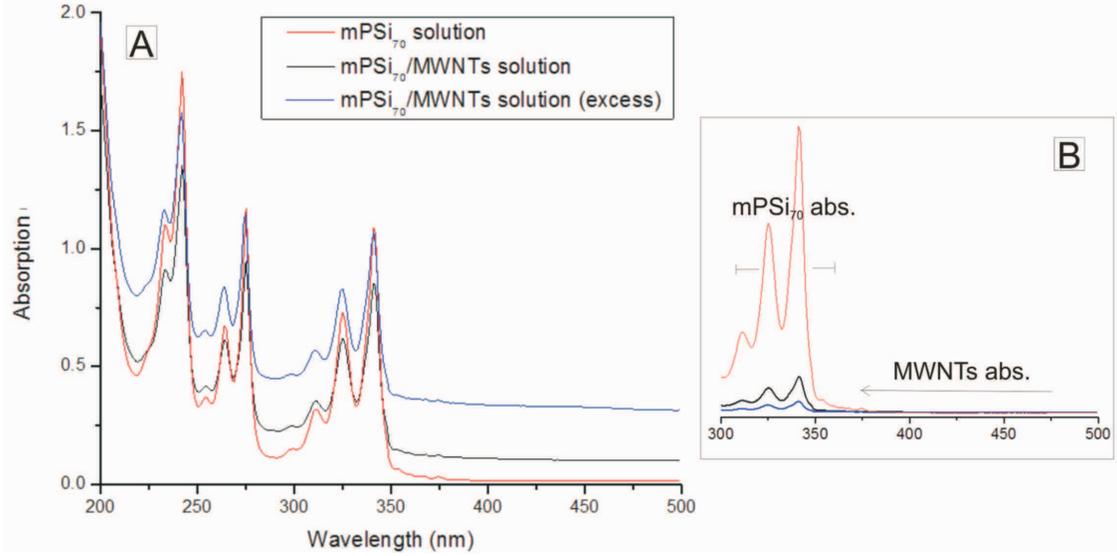}}
\caption{(A) Absorption spectra of PET solutions of mPSi$_{70}$,
the standard 4:1 w/w mPSi$_{70}$/MWNT, and the excess
MWNTs/mPSi$_{70}$; (B) the 300-500~nm  region selected to
emphasize the excess MWNTs absorption curve after normalization
against the standard mPSi$_{70}$/MWNT curve. All the absorption
spectra were acquired with [mPSi$_{70}$]$=5\times10^{-5}$~M in
PET. }
\label{mP70Si_abs}
\end{figure}

The direct interaction between the pyrene group and the CNT was
supported by the UV/Vis-absorption spectrum, Fig.~\ref{mP70Si_abs}.
It is generally known that MWNTs absorb
evenly across UV to near-IR, and the absorbance of pyrene is
suppressed when they are anchored to the
CNTs~\cite{Tanigaki07_PyreneCNT}. The mPSi$_{70}$ molecule was
shown to have similar absorption profile, although slightly
red-shifted compared to pure pyrene, with close to zero absorbance
in the 370-500~nm region. The complexation between MWNTs and
mPSi$_{70}$ (at the standard 4:1 w/w) introduced a steep rise in
absorption clearly identifiable in the $370$-$500$~nm region,
while the absorption peaks due to free pyrene groups decreased.
The anchoring of mPSi$_{70}$ molecules to MWNTs is better
demonstrated by comparing the spectra of the standard (4:1)
mPSi$_{70}$/MWNTs solution and the similar solution with an excess
amount of MWNTs. The mPSi$_{70}$/MWNTs (excess) solution was
prepared at the mPSi$_{70}$:MWNTs ratio of $\sim3:1$, while
keeping the overall surfactant concentration in PET constant
[mPSi$_{70}$]$=5\times10^{-5}$~M. By normalizing the absorption of
mPSi$_{70}$/MWNTs (excess) at the $370$-$500$~nm region against
that of the standard solution, one finds large proportional
decrease in the characteristic pyrene absorption bands, Fig.~\ref{mP70Si_abs}(B).
This means that the addition of extra
nanotubes had caused the mPSi$_{70}$ originally freely dispersed
in PET to attach to the additional tube surface. Clearly there
exists an equilibrium between the number of anchored mPSi$_{70}$
and the free surfactant molecules in solution.
Fluorescence spectroscopy (see Supporting Information) using an excitation wavelength of 300~nm
gives further evidence of the interaction between MWNTs and mPSi$_{70}$. The surfactant shows typical fluorescence characteristics of pyrene when dissolved in PET. However, the fluorescence was significantly suppressed in the the presence of MWNTs, which verifies the role of $\pi$-stacking due to the direct interaction of pyrene with MWNTs. Since free pyrenes are still present in the solution, the fluorescence is not quenched completely. This effect is different from what was observed for polysoap with pyrene side chains~\cite{Wang06_polysoap}, the fluorescence of which could be fully quenched by CNTs. However, the solubility of the CNTs enhanced by mPSi$_{70}$(10~mg/mL) is significantly higher than that achieved by polysoap surfactant (only about 0.075~mg/mL).

\begin{figure}
\centering
\resizebox{0.93\textwidth}{!}{\includegraphics{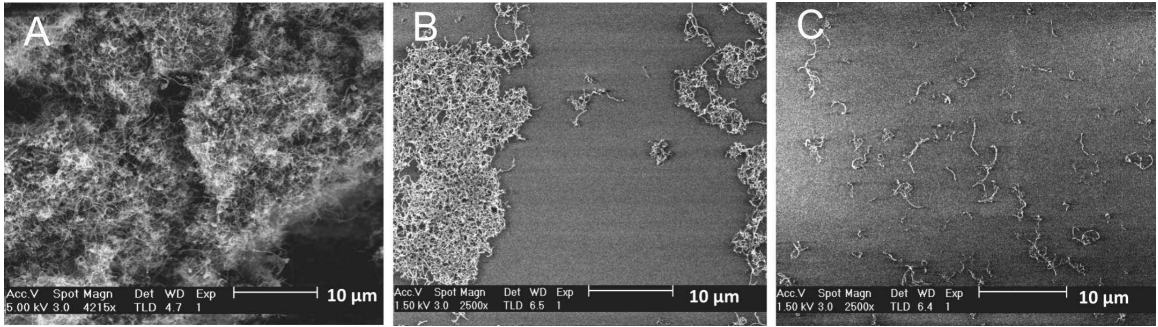}}
\caption{SEM images
showing: (A) as-received MWNTs; (B) clustered MWNTs after sonication in pure PET; (C)
well spaced, isolated MWNTs after sonication with mPSi$_{70}$.}
\label{MWNT_son}
\end{figure}

To visualize the final state of MWNTs dispersion achieved by
the combined effects of sonication and the surfactant,
SEM was applied. For the MWNTs reference sample
prepared in pure PET, large clusters of MWNTs were observed, Fig.~\ref{MWNT_son}(B).
In contrast, Fig.~\ref{MWNT_son}(C)
showing the dispersion assisted by mPSi$_{70}$ indicates that most
tubes were isolated and well-spaced even after toluene washing away
the remaining surfactant.
The length of the dispersed tubes has clearly been shortened
compared to the quoted 5-15~$\mu$m starting values. This is a side
effect of sonication which has been quantified in
\cite{Ahir08_CompositeReview}. With 1~hr sonication, most tubes
were getting close to the theoretical limiting length $L_{\rm
lim}$; this limiting length was calculated to be $\sim$2-5$~\mu$m
using the formula $ L_{\rm lim} = \sqrt{{d^2 \sigma^*}/{2\eta
(\dot{R}_{i}/R_i)}}$~\cite{Ahir08_CompositeReview}, given that the
nanotube diameter $d \sim$50-100~nm, the breaking strength
$\sigma^* \sim 4$~GPa for CVD MWNTs~\cite{Xie00_CNTStrength},
$\eta \sim 0.01$~Pa.s for typical low-viscosity solvents and
$\dot{R}_i/R_i \sim 10^8~\hbox{s}^{-1}$ for a cavitation bubble
implosion event during sonication (with $R_i$ and $\dot{R}_i$ being
the bubble radius and the bubble wall velocity respectively). Well-defined nanotube lengths
could only be determined in the surfactant-assisted dispersion in Fig.~\ref{MWNT_son}(C), whereas the clustered system of Fig.~\ref{MWNT_son}(B) results in spite of shortening of the tubes.

\subsection{PhrSi$_{60}$}
\begin{figure}
\centering
\resizebox{0.5\textwidth}{!}{\includegraphics{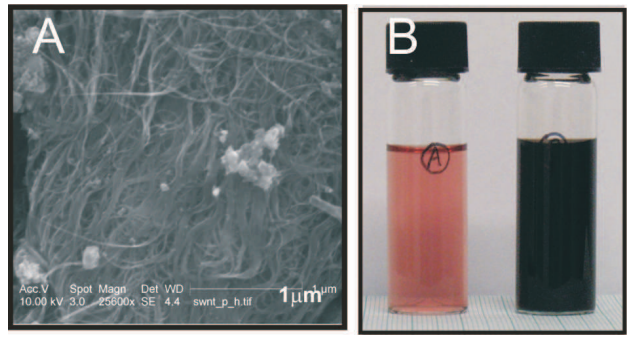}}
\caption{ (A) SEM of as-received SWNTs. (B) Bottled solutions with
pure PhrSi$_{60}$-PET on the left hand side, and
sonicated 0.15~mg/ml SWNTs-PET solution on the right hand side
(PhrSi$_{60}$:SWNTs$=$65:1 w/w). \ }
\label{SW_soln}
\end{figure}

The porphyrin-based surfactant PhrSi$_{60}$ was able to
effectively solubilize SWNTs but gave only negligible  solubility
enhancement for MWNTs in PET. The required surfactant to CNT ratio
was higher for the PhrSi$_{60}$ and SWNT combination; we found
that the samples with the ratio $\sim 65:1$ performed best. With
sonication treatment, a homogeneous solution of up to
$\sim$0.5~mg/ml in PET (solubility limit) could be obtained.
Figure~\ref{SW_soln}(B) illustrates a pink solution of pure
PhrSi$_{60}$ in PET, and a stable 0.15~mg/ml SWNTs solution with
PhrSi$_{60}$:SWNT$=65:1$ (w/w). Compared with the mPSi$_{70}$-MWNT
case in the previous section, the stability of SWNTs solution
achieved by the use of PhrSi$_{60}$ was not as strong. It was
found that the dark supernatant usually remained stable for about
two weeks after which precipitation started to appear, while for
mPSi$_{70}$/MWNTs, their PET solutions remained fully stable after
that time for similar concentration of CNTs.

\begin{figure}
\centering
\resizebox{0.99\textwidth}{!}{\includegraphics{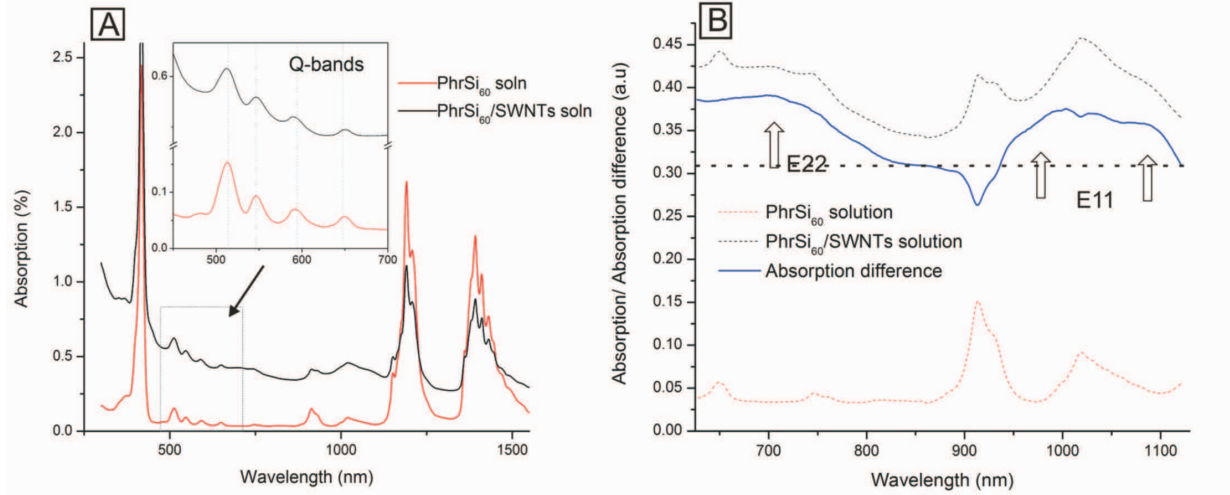}}
\caption{(A) absorption spectra of
PhrSi$_{60}$-PET solution, and that of the sonicated
PhrSi$_{60}$/SWNTs solution, with the insert emphasizes the absorption due to Q-band transitions;
(B) the difference between pure
PhrSi$_{60}$ and PhrSi$_{60}$/SWNT absorption with arrows
indicating the regions of absorption bands due to SWNTs. }
\label{PhrSi60_abs}
\end{figure}

We again employed absorption spectroscopy to investigate the
interaction  between the active conjugated group of the surfactant and the
CNTs, with results shown in Fig.~\ref{PhrSi60_abs}(A). Similar to
that of mPSi$_{70}$/MWNTs in PET, one finds an overall increase in
baseline absorption from visible to near-infrared wavelength,
indicating the general presence of dark objects (nanotubes in this
case) suspended in solution. In addition, the relative decrease in
the level of porphyrin absorption is also obvious, for example, in
the 1100-1250 and 1300-1500~nm bands. The peak position associated
with the Soret transition~\cite{Collman77_Soret,Lin94_Soret} of
PhrSi$_{60}$, i.e. the 416~nm band, remain unchanged. The Q-bands (see insert)
at 511~nm, 547~nm seemed to be unaffected by the addition of
SWNTs, however, the weaker Q-bands at 593~nm and 649~nm have
shifted to 588~nm and 652~nm, respectively. Since some free
PhrSi$_{60}$ molecules were still present in solution, the
absolute shift in the positions of these minor peaks is expected
to be greater after deconvolution. Theoretically, the
incorporation of SWNTs should induce additional absorption bands
due to the electronic transitions in SWNTs. Although these bands
are hard to distinguish in Fig.~\ref{PhrSi60_abs}(A), they can be
identified in Fig.~\ref{PhrSi60_abs}(B), where the difference in
absorption between the SWNT and the reference solution is plotted.
Since no obvious peaks are identified in the pure PhrSi$_{60}$ and
PhrSi$_{60}$/SWNTs solutions in the $\sim 850$~nm region, the
corresponding absorption difference was selected as the reference
level. If the level of absorption difference was above this level,
it means there is an extra absorption component in addition to the
uniform ``dark absorption'' and the absorption due to
PhrSi$_{60}$. Through this data treatment, we identify the unique
broad absorption band due to SWNTs at $\sim$950-1110~nm
(corresponding to transition centered at $\sim 1.2$~eV, E$_{11}$)
which is clearly a combination of several absorption peaks. A less
obvious band is located in the 700-770~nm region (as determined by
Peak Analyzer program, OriginLab) which corresponds to the SWNT
E$_{22}$ transition centered at $\sim 1.7$~eV.
The fluorescence of porphyrin was also quenched in the PhrSi$_{60}$-SWNTs complex (data in Supporting Information). Normalized emission spectra of PhrSi$_{60}$-SWNT overlaps with that of the pure PhrSi$_{60}$ in PET. We did not observe shifting of emissions which would occur if the electronic structure of porphyrin were perturbed. Just to note, Casey \textit{et al}~\cite{Casey07_phr_SWNT} also observed similar zero spectrum shift in their porphyrin-SWNT system. Further work is required in order to establish whether electron- or energy transfer could take place between the PhrSi$_{60}$ and SWNTs.

% E = 1240nm/rau eV
% molecular weight of Phr-60Si = 790 + 4*(70*(28+17*2+16))=22630 g/mol
% but Yan Ji gave me a value of 15,000 g/mol
% actual concentration used for spec testing is 0.028g/ml, which yields the
%molar used to be %1.87E-06 mol/ml that is 2*10^-3 mol/L.
%but purification wise is unknown therefore, we first assume it is 10% it is
%2x10^-4 molar, %because the peak absorption at 417nm is not that high
%compared to the 5x10^-5 molar of Phr-10C in PET. (NB. a different spectrometer
%was used %though!!)

\subsection{Composite fabrication}
\begin{figure}
\centering
\resizebox{0.7\textwidth}{!}{\includegraphics{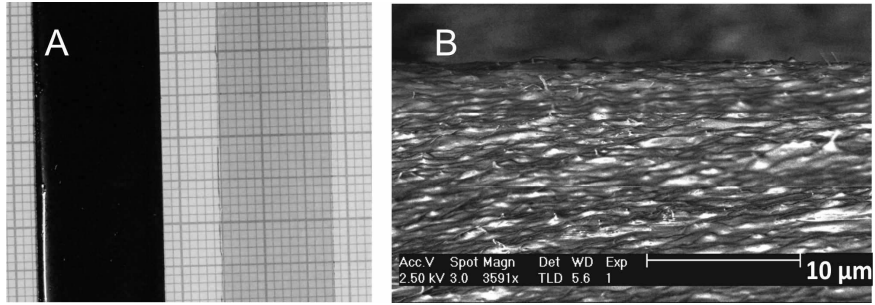}}
\caption{(A) Examples of 1~mm thick films of
mPSi$_{70}$/MWNTs-PDMS  composites elastomers on graph paper as
background: left-hand-side, 4~wt\% MWNT; and right-hand-side
0.06~wt\% MWNT. (B) SEM image showing the fractured surface of a
$4$~wt\% mPSi$_{70}$/MWNT-PDMS composite. } \label{MW_PDMS_comp}
\end{figure}

The special chemistry of the mPSi$_{70}$ and PhrSi$_{60}$ dangling
chains means that they are naturally compatible with PDMS elastomer
matrices. Although a growing number of applications ranging from
thermal/fire protections~\cite{Beigbeder08_CHPi} to transparent
conductive thin film~\cite{Zhang06_TransConducFilm}, have been
reported for CNT-PDMS composites, the issues associated with CNT
dispersion persist. With good stability even at high CNT
concentrations, our surfactants (mPSi$_{70}$ for MWNTs, and
PhrSi$_{60}$ for SWNTs) can bring two major advantages for the
processing of silicone nanocomposites. Firstly, much smaller
quantity of solvent is required during solution processing, as
opposed to the procedure of Giordani et al\cite{Giordani06_Dilution}. For example, pure PDMS
melt (Polydimethylsiloxane, Sylgard 184$^{TM}$ from Dow Corning)
 could be directly dissolved into the sonicated
surfactant/CNT-PET solution at a $\sim1:4$ PDMS to PET volume
ratio to produce a homogeneous composite. The second advantage is
that homogeneity in the bulk (3D) structure can be ensured, unlike
many cases when only homogeneous 2D film structures produced
because of sedimentation. In particular, we were able to
shear-align CNTs during casting with associated shear of the top
surface, which was then preserved by crosslinking of elastomer
matrix. Images of the finished products for higher and lower
bounds of concentrations of mPSi$_{70}$/MWNT-PDMS composites are
shown as an example in Fig.~\ref{MW_PDMS_comp}(A). The SEM image in Fig.~\ref{MW_PDMS_comp}(B)
illustrates isolated MWNTs distributed evenly across the fractured
surface of a $4$~wt\% surfactant-stabilized MWNT in PDMS composite.
Good interfacial adhesion is found between the nanotube and the
matrix. Further study is focused on testing the electronic and
photo-mechanical properties of these composites.

\subsection{Discussion}~\label{Discussion}
From the above results it is evident that direct and strong
interaction  between pyrene or porphyrin moiety and the corresponding
CNT surface exists despite one or multiple dangling siloxane chains, each
with lengths of $n\sim$60-70, attached to these active
centers. mPSi$_{70}$ and PhrSi$_{60}$ demonstrate interesting
selectivity towards the type of nanotubes they can disperse. Interaction
between pyrene and CNT surfaces is considered as $\pi-\pi$
stacking in nature~\cite{Chen02_PiStacking}, thus the local
curvature of CNT surface determines the optimum strength of such
an interaction because the pyrene group has a rigid planar
structure. This gives possible explanation to why the pyrene
derived surfactant, mPSi$_{70}$, did not greatly enhance the
solubilization of our SWNTs, whose local surface curvature was
too high. Although there are reports in the
literature~\cite{Chen02_PiStacking,Chen01_ProteinImmobil,Ehli06_PyrenePorphyrin}
claiming unbundling of SWNTs by some pyrene-derived molecules in
aqueous solutions, we would like to support the argument that
these molecules work better on SWNTs with larger
diameters~\cite{Ehli06_PyrenePorphyrin}. In future studies, it
will be interesting to compare the effectiveness of mPSi$_{70}$
for CNTs of a controlled range of diameters .
%
%xxxxxxx6:6 binding center for porphyrin~\cite{Hasobe06_PhrSWNTSolarCell}

However, the above explanation is not applicable to the porphyrin-based
surfactant PhrSi$_{60}$. If the entire porphyrin group is
seen as the active center, it would exhibit an even wider rigid
planar shape than the pyrene moiety. Nevertheless, the resulting
PhrSi$_{60}$ surfactant was highly effective for highly curved SWNTs, and
ineffective for a much more flat MWNT surface. It has been widely
suggested that the donor
electrons on the amine group (constituent in porphyrin)
could interact favorably with the CNT surface. This is a result of
the high curvature of the CNT graphene layers (sp${}^2$ bonds) which
renders them electrophilic, and therefore a favorable interaction can take
place between electron-accepting CNT surface and  a donor molecule via charge
transfer/sharing~\cite{Kong01_OrganicAmineGrafting,Sun01_anilineCNTinteraction}.
If these amine groups are the true active sites in the porphyrin
moiety, the comparative curvature argument utilized above is no
longer relevant. In addition, porphyrin molecules were also found to
selectively interact with semi-conducting SWNTs but not the
metallic ones~\cite{Chattopadhyay03_SepSemiMetalRaman}. In this
sense, one expects that about one-third of the initially dispersed
PhrSi$_{60}$/SWNTs were not stable in the solution since the
as-received SWNTs should have approximately 1:3 proportion of
metallic tubes. This may
explain the partial precipitation of SWNTs after two weeks' standing,
as well as the ineffectiveness of PhrSi$_{60}$ towards MWNTs (which are
all metallic in nature). If confirmed, this effect may offer a new way of separating
metallic from semiconducting SWNTs from a bulk sample.

Next we turn our attention to the roles played by the
dangling chains of our surfactants. From the above results,
it seems that the long
siloxane chain is acting much more effectively than the shorter
carbon chains.  Although molecular-dynamics simulations
\cite{Beigbeder08_CHPi} indicated that siloxane chains can
interact favorably with the CNT surfaces through CH-$\pi$
interaction, we did not find it the case in our experiments.
Firstly, our earlier rheological studies~\cite{Huang06_PRB}
revealed that re-aggregation would take place over time even for
the initially homogeneous dispersions of CNTs in PDMS. To further
prove the ``neutrality'' between neat siloxane chain and CNTs
surface experimentally, pure SWNTs and MWNTs were sonicated in two
low-viscosity grade silicone oils  (5~cst and 100~cst, Dow
Corning 200 Fluid series). The 100~cst silicone oil was suggested
to have a molecular weight of
$\sim$6000~g/mol~\cite{Povey02_SiliconeOilMr}, similar to that of
the siloxane chains attached to pyrene/porphyrin center in our
surfactants. Results showed that the sonicated MWNTs exhibited
near-complete sedimentation after one day standing in both media.
For SWNTs, aggregated clusters of tubes were visible but could
remain in suspension for weeks in the more viscous 100~cst silicone
oil. A more convincing (and relevant) fact is that we attempted to
disperse both kinds of our CNTs in the actual reactant, mono(hydroxyalkyl)
terminated PDMS (see the details of PhrSi${}_{60}$ synthesis above)
and have not detected any solubilization enhancement. We
thus conclude that the dangling siloxane chains do not interact
favorably with the CNT surfaces. However, as in classical
amphiphilic surfactants, once the active groups are attached to
the tubes,  the long flexible chains play an important role in
``suspending'' the tubes in solvent by providing the steric
repulsion preventing re-aggregation. Therefore, the length of
dangling chains is crucial to dispersion, and we will demonstrate
below why a chain length of $\sim$60-70 was adequate.

If one assumes the siloxane chain grafted to an anchored
pyrene/porphyrin group remains in solution taking up an
approximate Gaussian coil conformation, the radius of gyration
$R_{g}$ of this dangling part is estimated to be $\sim$1.3~nm.
This is calculated based on $R_{g} = \sqrt{n}\, a_{\rm SiO}$ with
the number of siloxane units $n \sim$60-70 and Si-O step length
$a_{\rm SiO} \sim 1.6$~{\AA}. (NB: unlike the ``wrapping''
mechanism suggested between SWNTs and free PDMS chains in the
molecular dynamics simulation of Beigbeder et al\cite{Beigbeder08_CHPi}, we
propose that the siloxane part should stay in the surrounding
organic solvent for our surfactant molecules with active centers).
Firstly, one can notice that the siloxane ``coil'' provides a
barrier just beyond the Van der Waals interaction distance
$\sim$1~nm~\cite{ShvartzmanCohen04_SelectiveDispersion}. Further
increase in siloxane chain length may decrease the mobility of the
surfactant molecules in solution, reducing the probability of them
attaching to the nanotube surface, and also impose lateral packing constraints for
the active centers on the surface.

Based on the above considerations we would like to propose that attachment
of  carbon chains of reasonably long length could also lead to
effective solubilization of SWNTs/MWNTs in an organic solvent, to
a comparable extent.  The main obstacle to test this suggestion
lies in the difficulty of synthesizing surfactants with a
sufficiently long (n$\sim$70) carbon chains.

\section{Conclusions}
Through separately studying the CNT-affine and solvent-affine
parts of a dispersant for non-polar hydrophobic solvents, we
conclude that the choice of an active center determines the type
of CNT to be dispersed, while the length of the dangling chain
determines the final stability of CNT in solution. According to
our results, mPSi$_{70}$ was found to effectively disperse MWNTs
in PET with solubility over 10~mg/mL, and PhrSi$_{60}$ could
enhance the dispersion of SWNTs in PET with a provisional
solubility limit of 0.5~mg/ml.  Overall, the relatively
straightforward process of organic synthesis for all these
surfactants make them readily scalable for industrial
applications.

%%%%%%%%%%%%%%%%%%%%%%%%%%%%%%%%%%%%%%%%%%%%%%%%%%%%%%%%%%%%%%%%%%%%%
\subsection*{Acknowledgement}

The authors thank T. Hasan and L. Payet
for help in spectroscopic measurement, R. Cornell and C.M. Amey for TGA measurements, and A. Ferrari, O. Trushkevych and M. Shaffer for useful discussions. This work has been supported by the EPSRC (EP/D04894X), ESA-ESTEC (18351/04), the
Gates Cambridge Trust and St John's Benefactor Scholarship. \\

\noindent Supporting Information Available for this article contains characterization data of thermogravimetric analysis (TGA) of CNTs, NMR and infrared spectra of our surfactants, and
 the fluorescence spectra of these surfactants on their own and in CNT suspension.
 This information is available free of charge via the Internet at http://pubs.acs.org/.

%%%%%%%%%%%%%%%%%%%%%%%%%%%%%%%%%%%%%%%%%%%%%%%%%%%%%%%%%%%%%%%%%%%%%
%\bibliography{achemso}
%

\newpage
\begin{center}\textbf{\large Supplementary Information}\end{center}

\begin{figure}[h!]
\centering
\resizebox{0.7\textwidth}{!}{\includegraphics{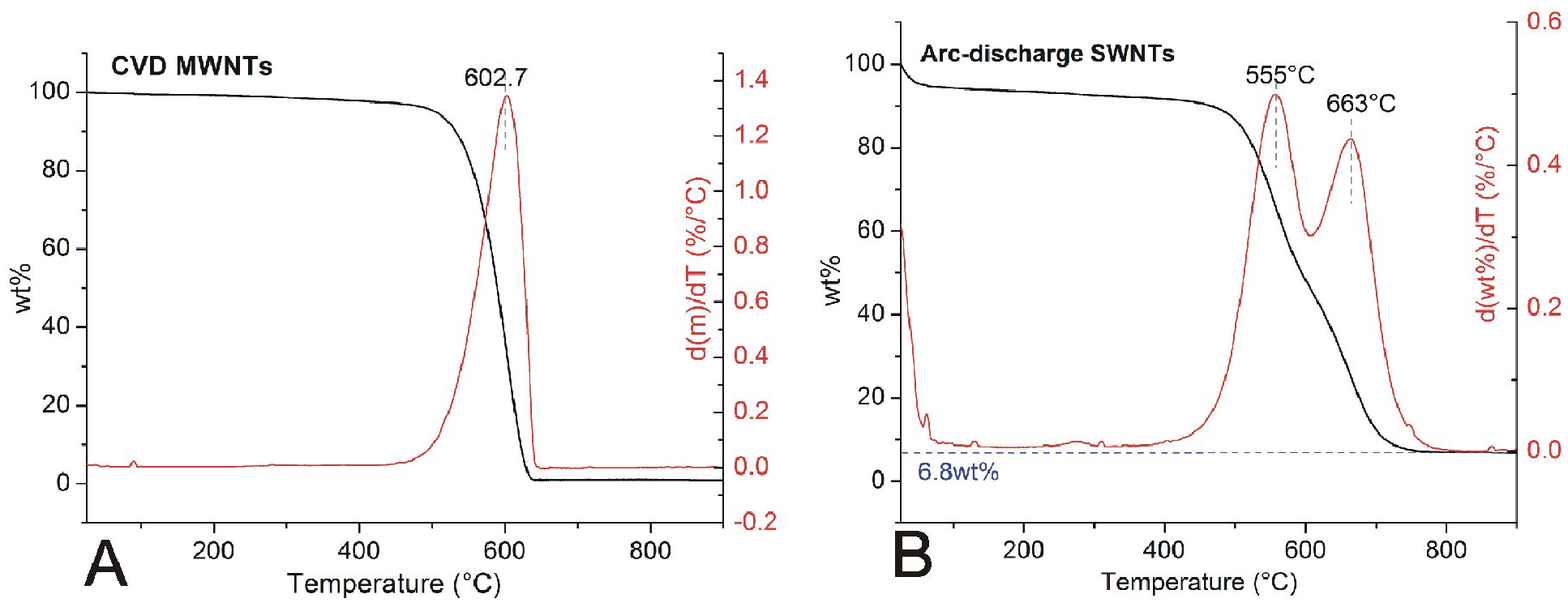}}
\caption{(A) Thermo-gravimetric analysis (TGA) of MWNTs; (B) TGA of SWNTs} \label{s1}
\end{figure}

\begin{figure}[h!]
\centering
\resizebox{0.7\textwidth}{!}{\includegraphics{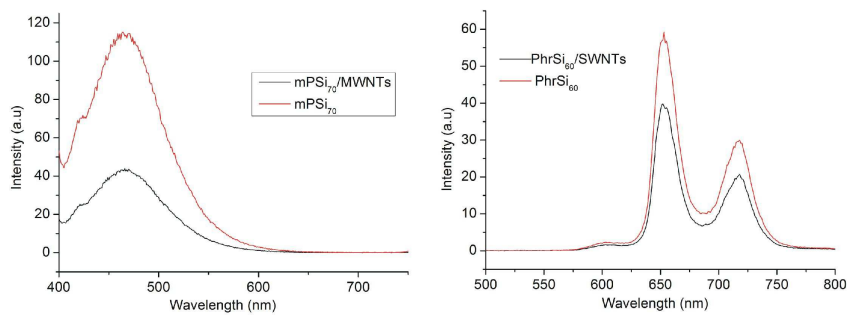}}
\caption{Left panel: plot showing fluorescence spectra of mPSi$_{70}$-PET reference solution, and mPSi$_{70}$/MWNT-PET solution;
Right panel: plot showing fluorescence spectra of PhrSi$_{60}$-PET reference solution, and PhrSi$_{60}$/MWNT-PET solution } \label{s5}
\end{figure}\

\end{document}